# The Mad Hatter's Cocktail Party: A Social Mobile Audio Space Supporting Multiple Simultaneous Conversations

Paul M. Aoki, Matthew Romaine[†], Margaret H. Szymanski,
James D. Thornton, Daniel Wilson[*] and Allison Woodruff

| Palo Alto Research Center | [†]CCRMA, Dept. of Music | [*]Robotics Institute |
|---|---|---|
| 3333 Coyote Hill Road | Stanford University | Carnegie Mellon University |
| Palo Alto, CA 94304-1314 USA | Stanford, CA 94305-8180 USA | Pittsburgh, PA 15213-3890 USA |

**ABSTRACT**

This paper presents a mobile audio space intended for use by gelled social groups. In face-to-face interactions in such social groups, conversational floors change frequently, e.g., two participants split off to form a new conversational floor, a participant moves from one conversational floor to another, etc. To date, audio spaces have provided little support for such dynamic regroupings of participants, either requiring that the participants explicitly specify with whom they wish to talk or simply presenting all participants as though they are in a single floor. By contrast, the audio space described here monitors participant behavior to identify conversational floors as they emerge. The system dynamically modifies the audio delivered to each participant to enhance the salience of the participants with whom they are currently conversing. We report a user study of the system, focusing on conversation analytic results.

## INTRODUCTION

The work reported in this paper is motivated by the following two observations. The first is technological: incremental improvements in wireless networking (both high-bandwidth digital cellular networks and wireless local-area networks), semiconductors and power storage technology are coming closer to enabling pervasive, continuous, near-toll-quality audio communication services to mobile users. The second is social: as wireless communication becomes pervasive, changes arise in social group structure and communication accountability. That is, as wireless technology makes possible increased availability for communication, the desire for and *expectation* of availability also tend to increase between members of gelled social groups (see, e.g., [10,11]).

One recent manifestation of these effects has been an increased use of existing services, such as mobile telephony. However, one can also ask what new services are becoming possible that might address needs that are unmet by today's services. In the extreme, the observations suggest that it is reasonable to revisit the last decade's research on media spaces [3,7], which provide lightweight communication and awareness information through the use of open, multi-party channels. Specifically, they suggest that a mobile, audio-only media space might be suitable as a tool for binding together gelled social groups.

Building such a tool raises a large number of difficult social questions and design problems, and these are the subject of our ongoing research program. Some are long-standing media space issues, such as privacy. However, some problems are entirely new; some of the most interesting of these concern the actual facilitation of within-group social communication. We address one such problem here.

In this paper, we describe a system that is intended to facilitate lightweight group discussion within an audio space. From the literature and our own design ethnography [27], we observed that the highly dynamic structure of social group conversations was poorly served by existing audio communication systems. Although schisming – the transformation of one conversational floor into two simultaneous conversational floors – is common in such conversations [9], it is addressed with very heavyweight mechanisms in existing systems (if it is addressed at all). Such mechanisms seem poorly suited for mobile use. Drawing on conversation analysis [16], we enhanced our audio space system with a machine learning component that analyzes participant turn-taking behavior to identify conversational floors as they emerge, noting which participants are in which floor. The system dynamically modifies the audio delivered to each participant to enhance the salience of the participants with whom they are currently conversing and to reduce the salience of the participants with whom they are not currently conversing. Each participant therefore receives a customized mix of all floors, tailored to their current conversational status.

---

[†] Current address: Sony BA Laboratories, 6-7-35 Kitashinagawa, Shinagawa-ku, Tokyo 141-0001 Japan.

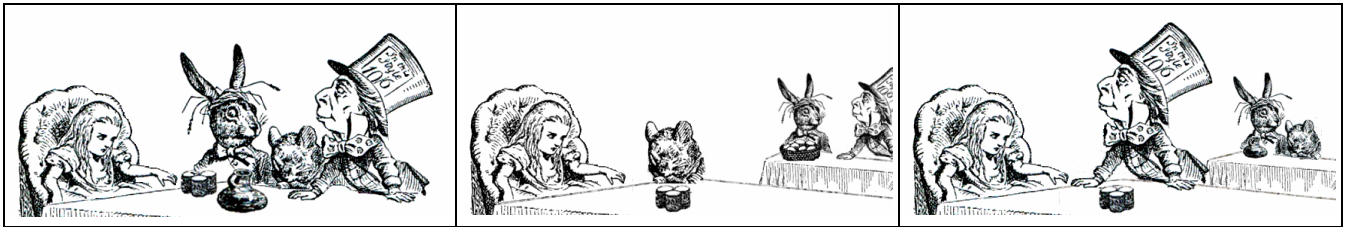

**(a) Participant A uses selective listening to attend to the desired floor.**  **(b) As floors reconfigure, the speakers not participating in A's floor appear to distance themselves spontaneously to assist her in floor separation.**

Figure 1. The cocktail party effect (left) vs. the Mad Hatter's cocktail party (right).[1]

We report on a comparative study in which users undertook highly concurrent social conversation in a conventional audio space as well as our prototype audio space. The results draw upon semi-structured user interviews and an applied conversation analytic study. The present prototype is moderately successful in its predictions but does make incorrect inferences. We therefore present two main categories of comparative results: an analysis of user conversation while the system operated as intended, and a description of the repair strategies used by participants when encountering problems. These findings have value that extends beyond the specifics of our own system, both in terms of research on human interaction and in terms of interactive system design.

The contributions of this paper are twofold. First, we present a novel audio space system that dynamically and automatically identifies conversational floors and modifies the audio heard by each participant so that they can hear their co-conversants more easily. Second, we present a conversation analytic study that (1) illustrates the types of highly problematic behaviors that arise in conventional audio spaces when participants attempt to carry on multiple simultaneous conversations, and (2) illustrates that automatic audio enhancements can effectively facilitate multiple simultaneous conversations.

The remainder of the paper is organized as follows. We first provide some background in various aspects of conversation. We then give details of our prototype. After describing the design of our user study, we present findings from user interviews and the applied conversation analytic study. We then turn to related work and conclude.

**BACKGROUND**
Before describing our system, we discuss three background topics. Each motivates a different aspect of the system, and each draws on a different discipline in the social sciences.

**On Design Challenges of Social Talk**
Many audio computer-mediated communication (CMC) systems are designed to support some variation of "informal workplace communication" [25]. Naturally, "friendly" social interactions often take place in the workplace, and interactions at work share many characteristics of social interactions outside of work (see, e.g., [1,7]). Hence, systems of this kind are studied and evaluated with some sensitivity to these issues.

Nevertheless, there are aspects of "friendly" social interaction which previous systems have never been designed to support. First, such interaction is often collaborative [8] and concurrent [5]. High utterance spontaneity – talk as "jam session," which is common in female social talk [5] – is particularly at odds with the strict turn-taking behavior that works best in an audio conference. Second, social group interactions often have highly dynamic floor structure, which we have seen in the literature on face-to-face interaction [9] as well as our own design ethnography of students using lightweight, long-range audio communication systems [27].

In a highly fluid conversation, it would be very tedious to have to designate each recipient explicitly, whether in the form of a user interface gesture or a verbal address ("Alice, Bob, Charlie, do you think…"). Further, explicit user interface gestures are likely to be problematic for mobile users. We have taken the following as a design challenge: the use of natural conversational behavior as an implicit interface for floor management in place of reliance on an explicit mechanism.

**On Cocktail Parties and Conversational Enhancement**
Listeners often monitor multiple audio sources. In the case of multi-party conversation, such monitoring provides both activity awareness ("Oh, Bill stopped talking to her, let's go see him") and interaction resources ("Hey, I heard you mention my favorite book…").

The term "cocktail party effect" generally "refers to the difficulty the listener has in following one voice in a mixture of conversations" [4], i.e., to the human capacity to separate sound sources using selective attention in listening (Figure 1(a)). Selective listening requires significant concentration for humans and is not a solved problem for computers. Furthermore, in face-to-face conversation, spatial (directional) hearing is a major facilitator of human selective listening. Some headphone-based multi-party systems attempt to simulate this by, e.g., spatializing the audio channels. Systems that cannot provide a controlled stereo environment (such as a conventional speakerphone) necessarily require listeners to separate sources without such information, which is far more difficult.

---
[1] The reference is to "A Mad Tea-Party" in Lewis Carroll's *Alice's Adventures in Wonderland,* in which the characters continually change places for the benefit of a single participant. Figure 1 is based on Tenniel's illustrations (obtained from Project Gutenberg, promo.net/pg).

Our approach to conversational enhancement sidesteps most of these issues. In "The Mad Hatter's Cocktail Party," the speakers with whom you are not interacting seem to move far away from you, i.e., their audio attenuates. Such attenuation makes it much easier to understand the speaker of interest while maintaining the ability to monitor other speakers (Figure 1(b)). This happens automatically, unlike in other systems which support attenuation using explicit interface gestures (e.g., [14]). Spatial audio is unnecessary because a gross difference in volume makes it easy to listen to the speaker of interest while preserving the ability to monitor other speakers.

### Relevant Concepts from Conversation Analysis

We briefly review some of the terminology and findings that underlie the design of the system and our observational analyses. Both rely on the research and methods of *conversation analysis* [16], a methodology for analyzing how human interaction is organized into sequences of action.

The organization of taking turns at talk is fundamental to conversation. One of the ways in which turn-taking organization operates is by specifying opportunities for *speaker change* at *turn-constructional units* (TCUs) from which turns at talk are composed [17]. This enables listeners to monitor and project the completion of others' TCUs in order to time the initiation of their own turns properly. Completion of a TCU is often accompanied by a pause in speech, making a *transition-relevance place* (TRP) where speaker change may occur. Research has characterized the duration of such pauses in two-party conversation [23,26] and has also indicated that sustained periods of *simultaneous speech* are infrequent [17]. (When overlap does occur, one or more of the overlapping speakers typically drop out.) When trouble arises in speaking, hearing or understanding, participants can initiate *repair*, an organized set of methods and practices designed to keep talk going [19].

Multi-party conversations may consist of a single *floor* in which participants orient to each others' turn-taking behavior as just described. However, in casual multi-party conversation, a given floor frequently *schisms* into multiple floors [9]; this process can be quite dynamic, in that floors may merge and new schisms can occur.

A key observation is that when two simultaneous conversational floors are on-going, participants in one do not orient to the turn-taking organization of the other conversation. This has two implications: (1) speakers in one conversation no longer align the initiation of their TCUs with the TRPs of the other conversation, and (2) speakers in the distinct conversations overlap their talk much more than if they were participating in a single floor. These implications form the basis for our current machine learning algorithm, which is described in the next section.

### PROTOTYPE

In this section, we describe the design and implementation of the prototype mobile audio space employed in our user study. We first describe our audio processing and delivery infrastructure. We then discuss the floor assignment subsystem, which contains the machine learning algorithm.

### Audio Infrastructure

The system consists of two main components: a client audio tool running on a handheld computer and a server application running on a desktop workstation. A wireless network is used for client access and a conventional switched network connects the wireless access point to the rest of the system infrastructure.

The client is a modified version of UCL's RAT software (www-mice.cs.ucl.ac.uk) on an HP/Compaq iPAQ™ H3650 running Linux. In the study reported here, the client was simply used as a transceiver for the full-duplex audio space implemented by the server. We used conventional single-ear telephony headsets with boom microphones (requiring a small hardware modification to the iPAQ).

The server is an application written against the GStreamer (www.gstreamer.net) component framework on a PC running Linux. The server receives audio streams from each client, analyzes them, applies audio effects and performs mixing, and then transmits audio streams back to each client.

In this study, we made conservative assumptions about the available network capabilities. The clients communicated with the server using RTP over an IEEE 802.11b wireless local area network (WLAN), transmitting audio coded at toll-quality (64kb/s) data rates. We measured end-to-end latency (microphone to earphone) at 350ms. Hence, the results reported should be applicable to existing telecommunications infrastructure as well as to WLANs.

### Floor Assignment Subsystem

The floor assignment subsystem, which runs within the server application, continually determines the most likely configuration for the conversational floor(s) from the content of the audio streams. It runs a voice activity detector on each incoming audio stream, producing binary classifications (speech/non-speech) at 1ms intervals. It then extracts temporal features from each pair of binary streams and applies a Naïve Bayes learning algorithm to those features. The learner outputs probabilities of mutual floor participation for each pair of users; it considers these probabilities in view of a model of valid floor configurations, choosing the most likely configuration.

*Offline training and feature set*. The training data was based on data captured from face-to-face social conversation. We made individual recordings of eight participants in copresent English conversation. The participants consisted of three female and five male employees of a research center, all between the ages of 20 and 40 – three native U.S. English speakers, two native non-U.S. English speakers, and three non-native speakers from Europe and Asia. The analyzed portion of the conversation contained 1849 turns over 53 minutes. The conversation contained eleven identifiable floors, of which as many as three were active at once. Afterward, the audio was segmented into turns, and each turn was labeled with a floor assignment by the team's conversation analyst.

For each pair of speaker segmentations, we automatically extracted two temporal features: TRP positioning and simultaneous speech. TRP positioning for speakers A and B is the distance between the starting endpoint of A's most recent utterance and the final endpoint of B's utterance that mostly closely precedes A's utterance. Simultaneous speech is the amount of overlap between A's utterances and B's utterances during a given time period; simultaneous speech was broken down into three subfeatures, one for each of three disjoint time periods: 0s-1s, 1s-14s and 15s-30s. The learner was trained using these features and the floor labelling. The resulting parameters were used to generate a set of distribution lookup tables. Each table entry constitutes a likelihood that a given feature value supports the hypothesis described by the distribution (either that a pair of speakers are participating in the same floor or are not). These likelihoods are combined into a final posterior probability for each class.

*Online floor assignment algorithm.* While the system is running, the same features described above are extracted from the incoming audio streams. During each floor evaluation period (currently 30ms), the learner uses the features and the distribution lookup tables to produce a set of pairwise posterior probabilities that users are participating in the same floor. (A time correction is applied to account for network delays.) We then enumerate the valid configurations of disjoint subsets of users. In each configuration, the subsets form a weighted graph in which users are nodes and the posteriors are the edge weights. We choose the configuration with the highest mean edge weight as the most likely floor membership.

*Result of floor assignment.* At the end of each floor assignment period, the chosen floor configuration is used to set how speaker A's audio is presented to speaker B. In the current system, if A and B are participating in the same floor, then they hear each other at a "normal" volume level. Otherwise, they hear each other at a "quiet" volume level (currently 20% of the "normal" level). As a result, every participant in the audio space hears a customized mix.

**STUDY**

The study was exploratory and had two main purposes. First, we wanted people to use the system and share their impressions on topics such as utility and usability. Second, we wished to assess the system in an environment in which it would be (relatively) straightforward to perform a qualitative analysis of the conversational dynamics. For example, it would clearly be important for floor schisms to occur frequently.

*Participants.* The study participants consisted of three female and five male employees of a research center, again between the ages of 20 and 40 – six native U.S. English speakers, one native non-U.S. English speaker and one non-native English speaker. None of the participants were members of the group from which we had collected training data. The participants were organized into two sessions of four, each with roughly balanced sex distributions.

*Procedure.* Participants in each session were placed in distant locations and used two different systems to communicate. In the first, which we will call the *conventional audio space*, all participants could be heard at all times at the same volume without any dynamic volume adjustment. In the second, which we will call *Mad Hatter*, volume adjustments were applied as described in the preceding section. (The first system was actually the same as the second, with the volume adjustment disabled.)

We structured the participants' activity using a party game designed to stimulate spirited conversation. The game required the participants to pair into teams, form a team consensus on a given controversial question of opinion (each team had a different question), form a new consensus on how "most people" would answer the same given question, and then return to the main floor to find a new partner and repeat the activity. Teams were explicitly instructed to work on their questions simultaneously. Each round took approximately three minutes on average.

Participants were told they would be using two different systems, but they were given no instructions about the nature of these systems or how they differed from each other. Participants used the conventional audio space for approximately fifteen minutes, and then we interviewed them briefly on their impressions of this system. Participants then used Mad Hatter for approximately half an hour, and were interviewed for approximately half an hour about their impressions of this system and how it compared to the conventional audio space. The asymmetry of the procedure was intentional, motivated by considerations such as our concern that participants would not be able to maintain simultaneous conversations for a prolonged period of time in the conventional audio space. Because of the exploratory nature of our study and the type of analysis, this was appropriate; however, a controlled laboratory experiment would plainly require a different procedure.

*Analysis.* The analysis reported in the next section is based on transcription, summarization and clustering of the recorded semi-structured interviews. This data is used to capture subjective opinions and self-reported behavior.

The analysis in the subsequent section relies on conversation analytic methods. A key goal of conversation analysis is to examine social interaction to reveal organized patterns or practices. In this case, identifying these systematic practices required examination of the recorded audio space sessions (which are different for each of the participants) and the production of detailed transcripts for segments of interest.

**INTERVIEW RESULTS**

We now turn to selected findings from our semi-structured interviews. The main results reported in this paper, contained in the next section, derive from analyses of the structure of talk in the two audio spaces; the findings reported in this section provide helpful context for these analyses. In addition, the participants' comments provide several important design-relevant insights. We describe two categories of findings, one focused on reactions to various aspects of the system related to floor separation, and the other focused on self-reported use strategies.

**Floors: Desirability of Separation**

It is not obvious *a priori* that floor-based volume changes are a good idea, let alone automatic floor assignment. In fact, participants did respond positively to the general approach, and to the automatic approach to the degree that it worked. One particular failure mode was especially disliked, which has important design implications.

*Effect of floor-based volume changes.* In the conventional audio space, participants felt that it was difficult to conduct a conversation at the same time as the other team. For example, participants said it felt rude to speak over other people. Participants were also very sensitive to the fact that some voices (e.g., female voices) could be heard more easily than others.

By contrast, participants said that when Mad Hatter made the correct choice about floors, it was much easier to have multiple simultaneous conversations than in the conventional audio space. They reported that the volume manipulations were an effective technique for separating multiple floors, saying for example, "That was better because you could, you could focus on the other person a lot more easily. You knew there was another conversation going on, but it didn't distract you as much." Participants also said that when Mad Hatter assigned floors correctly, they could better hear other participants whose voices did not come across clearly in the conventional audio space.

*Effect of (in)accuracy.* Participants reported that when Mad Hatter made incorrect choices, it could be more difficult to use than the conventional audio space. Not surprisingly, they said that hearing a team member at a low volume was frustrating, as was hearing a non-team member at a high volume. Voices fading in and out were annoying when participants were attempting to converse – participants found it particularly frustrating when they were in a correct floor and then the system drew an incorrect inference and their team member faded away. Much of the frustration arose because participants did not feel they were in control of the system and they did not feel they could predict how the system would behave.

*Opinions on the current prototype.* Participants all felt that when Mad Hatter put them in the correct floors, it was preferable to the conventional audio space. They said it was soothing and easier to converse when they could hear the other participants at a reduced volume, as compared with the conventional audio space in which all participants were always at the same volume. Some participants felt that the conventional audio space was preferable to the current implementation of Mad Hatter, due the latter's current level of accuracy; however most of these participants said they would prefer Mad Hatter if it had modifications such as a "hold" button (meaning a button that they could press to remain in a given conversational configuration once Mad Hatter had chosen it for them).

**Strategies: Coping with Difficulty**

Participants reported using very different strategies in the two systems. Since an automatic system will never be completely accurate, an understanding of the ways in which users intuitively cope with difficulty will be relevant to any such system.

In the conventional audio space, participants described both listening and speaking strategies. The main listening strategy was tuning out non-team-members and focusing on their partner (i.e., selective listening using voice characteristics). While speaking, participants tried speaking more loudly, speaking more slowly, and delaying speaking until there was space in the conversation.

In Mad Hatter, strategies were generally related to speaking, often focusing less on the listener *per se* than on trying to get Mad Hatter to assign the desired conversation floors and to preserve the correct floors once they occurred. For example, participants tried to make short statements or take turns talking back and forth. Many of these strategies were used intermittently to get to the correct state; by contrast, the strategies in the conventional audio space were necessarily being executed at all times.

**CONVERSATION ANALYSIS**

In this section, we present findings obtained through our applied conversation analytic study. As in the previous section, we discuss two main topics of interest, which again center on floors and strategies. First, we examine floor-related action in the respective audio spaces and compare them to what we know about multi-party, face-to-face conversation (e.g., [8,9]). Second, we describe some means by which participants used repair practices in the face of communication difficulty. (Here, we discuss moment-by-moment behaviors as opposed to top-down strategies.)

Our analyses are based upon a collection of transcribed excerpts. Table 1 summarizes the notation used.

| X: | Participant X is speaking |
|---|---|
| Y: | *Participant Y is speaking in a different floor* |
| (*n*) (.) | *n* second pause; micropause |
| My **[talk ]**<br>   **[your]** talk | Alignment of overlapping speech or actions |
| a**:**  a | Elongated vowel; stressed speech |
| , | Falling intonation |

**Table 1.  Summary of transcription notation.**

**Floors: Orientation to Two Floors vs. One Floor**

In face-to-face conversation, in the steady state, it is axiomatic that speakers in different floors do not orient their utterances to each other. Participants adhered to this behavior when Mad Hatter configured the floors correctly, but they did not follow this behavior in the conventional audio space.

In the conventional audio space, participants frequently positioned their conversational turns relative to both floors (the floor of their own team and the floor of the other team). This interleaving of turns reduced overlapping talk between the two floors, which improved intelligibility at the cost of disrupting the rhythm of the individual floors. We present examples of two phenomena that illustrate this point.[2]

First, participants would often wait to speak until someone from another floor finished speaking.

**Excerpt I.**

| 1  | Y: | would you rather: (.) have to wear big clown shoes, |
| 2  | K: | four |
| 3  |    | ounces isn't that- |
| 4  |    | isn't [*that much, I  wo*]uld say liquid detergent, |
| 5  | B: | [big w<u>h</u>at shoes?] |
| 6  | Y: | big clown |
| 7  |    | shoes, (.) you know the |
| 8  |    | big [clumsy shoes  ] that a clown wears, |
| 9  | A: | [*what was that?*] |
| 10 | K: | I don't know, I'll |
| 11 |    | let you choose that one,[*since I chose the first*] one, |
| 12 | B: | [okay, I would rather:  ] |

In lines 1-12 of Excerpt I, we see that Y and K are essentially taking turns speaking – as soon as Y completes his turn in line 1, K follows immediately with his turn in line 2; as soon as K completes his turn in line 4, Y follows immediately with his turn in line 6, etc. The key point is that although K and Y are in "different" conversations on different topics (Y is paired with B, and K is paired with A), in this excerpt they are plainly following turn-taking patterns that would be appropriate for a single floor. B is also orienting somewhat to the conversation in which she is not a participant, positioning her overlapping turns with K as though they are in the same floor, e.g., in line 12, she initiates her turn at a TRP in K's turn.

Second, if someone from another floor were speaking, participants would employ strategies such as elongating a word's pronunciation to delay delivering important content until the other person stopped speaking.

**Excerpt II.**

| 1 | A: | [*liquid detergent then.*] |
| 2 | Y: | [so   my   question  i:s] would you rather… |

In line 1 of Excerpt II, A is just completing a response. Y allows the *pro forma* part of his question to overlap with A in line 2. However, he draws out the word "is" until A has finished, which allows him to continue with the question itself without speaking in overlap.

When Mad Hatter configured the floors correctly, participants generally positioned their talk relative to their own floor without regard to the conversation in the other floor. (We do not provide an example as this is exactly the behavior one would expect in face-to-face conversation.)

We can summarize this subsection as follows. In an environment with a single, static volume level, the participants frequently positioned utterances relative to multiple floors, a highly unnatural behavior. In an environment in which volume levels were adjusted to deemphasize speakers from other floors, participants exhibited natural turn-taking practices within each floor.

**Strategies: Failure and Repair**

Of course, highly problematic cases occurred in both spaces. The different nature of these communication failures led to different repair strategies.

In the conventional audio space, participants would often have difficulty hearing each other at all. They would attempt to initiate repair of misunderstandings with statements such as "I didn't hear the second question at all." Repair was generally more explicit, included more elements, and indicated more complicated trouble than commonly occurs in face-to-face conversation. For example, at one point, B said to her partner Y:

**Excerpt III.**

| 1 | B: | sorry, slow down, |
| 2 |    | I- I can't hear you eh- too well, |
| 3 |    | would you rather: go around carrying the w<u>h</u>at? |
| 4 |    | CPR dummy? |

This multi-element turn indicates, among other things, that B is having trouble hearing Y generally (hence the instructions for him to slow down in the future), that she has had trouble hearing this particular utterance, and that she is having trouble hearing or understanding the final noun phrase.

Because communication was generally so problematic in the conventional audio space, participants rarely made comments that were not related to their questions or to repair. Further, when one team had completed a question and was waiting for the other team to finish, they would typically remain silent or participate in the other team's discussion.

By contrast, when Mad Hatter configured the floors correctly, participants had little or no trouble communicating. Consequently, they rarely initiated repair. Similarly, participants were able to bridge transitions between floors quite smoothly and unremarkably as the floor assignment algorithm merged them into a single floor and split them apart again.

Because communication with Mad Hatter was easier than in the conventional audio space, participants would often make off-topic comments. For example, when a team had completed a question and was waiting for the other team to finish, the team that had completed their question would sometimes have a casual conversation in which they would discuss topics such as soccer. This phenomenon occurred in both sessions, and participants specifically said this type of conversation was enabled by the fact that the other team's conversation was at a reduced volume.

---

[2] In these excerpts, A and K are completing the following question: "Would you rather drink four ounces of liquid detergent or eat a tablespoon of dishwasher powder?" B and Y are answering the question: "Would you rather have to walk around carrying a CPR dummy or have to walk around wearing big clown shoes?"

Recall that in the interviews, participants found it particularly disturbing when Mad Hatter assigned them to the correct floor and then their partner faded away, replaced by the voice of a participant from the other team. Our analysis of these interactions provides insight as to why this case is so problematic.

Participants frequently take quite a while to detect that their partner has incorrectly been assigned to another floor. Consider a hypothetical co-conversant pair, A and B, and assume that A is speaking. In face-to-face conversation, B would quickly alert A to any communication difficulties. However, when Mad Hatter assigns participants to the incorrect floors, the reduced volume between A and B actually interferes with and delays this alerting process. Some sources of delay include: B may not be aware that A is speaking at all (e.g., if B can only hear participants in another floor); B may make utterances targeted at A, but A may not hear them (e.g., if A hears other participants talking loudly at the same time that B is talking softly); or B may remain silent until the floor assignment algorithm has corrected itself. When speaker A is unaware that there is a problem, our analysis indicates that A may continue for quite awhile. If repair is delayed for several utterances, both A and B will have difficulty in locating the point in the conversation at which the trouble originated. This requires them to do extra work to locate the source of the trouble – work that makes repair sequences more complex and heavyweight than would normally occur face-to-face.

In summary, both static volume levels and incorrect dynamic volume levels can result in situations that require complex and extended forms of conversational repair. Correct dynamic volume levels do not require this work, and our analysis suggests that users can usefully be given appropriate means of detecting and correcting key problematic situations. Perhaps more importantly, we saw that participants spontaneously began talking about "off-task" topics in the middle of the study – a strong indication of the naturalness of the Mad Hatter audio environment.

**RELATED WORK**

*Audio-based media spaces.* Relatively few true audio spaces have been developed, and of these, all have been on the desktop (e.g., the Interval systems [1]). The only mobile "audio space" was actually receive-only (Nomadic Radio [18]); other mobile systems are either push-to-talk or use a call-setup model – e.g., the commercial Vocera system (www.vocera.com) uses a wireless "handset" that accepts voice commands. Similarly, voice command and control systems (such as the radio/interphone systems used in military, aviation and spaceflight – e.g., [13]), are often used for extended communication within large groups but their "floor control" exploits a radio-derived interaction design and rigid operating procedures. Our system is the only true multi-party mobile audio space of which we are aware, and certainly the only one with automatic floor assignment.

*Side and parallel conversations.* Various multi-party audio systems have attempted to support "side" or "parallel" conversations. These fall into three main classes. First, some audio space systems modify the audio to improve speaker separation. Spatial audio helps with selective listening and is often used in this context (e.g., [6,18,21]) but its effectiveness has not been demonstrated for casual users in parallel conversations [21]. Second, some space-like systems provide mechanisms by which participants can be temporarily partitioned. Commercial teleconferencing systems provide heavyweight partitions (subconferences). Some research systems have supported more lightweight partitioning mechanisms, albeit ones that still used explicit user interface gestures (e.g., selective inclusion by the speaker, as in "whispering" [2]). Third, some systems that are not space-like resemble a kind of audio chat (e.g., Impromptu's chat mechanism [20] and simplex push-to-talk systems [22]). Chat serializes utterances (generally whole turns), which increases the intelligibility of "parallel" utterances but also interleaves them in a serial stream. Our system uses dynamic, automatic floor assignment to manage parallel conversations.

*Dialogue formation.* Speech technology research addresses many loosely related problems, which include scene analysis; source separation and classification; and segmentation by genre, topic and speaker. The few systems that consider multi-party interactions (which include multi-agent interactive environments [24] and meeting-room capture systems [12]) seem to address specific problems that do not require them to perform real-time analysis of multi-party floor participation by humans. The system most closely related to ours is BBN's Gister, which links utterances in an air traffic control radio channel into controller/pilot dialogues [15]. Gister is able to exploit simple *turn adjacency*, which is a meaningful concept in simplex radio communication (a domain that is highly structured by convention – simultaneous speech and schisms essentially never happen). Adjacency becomes less meaningful in true multi-party conversation because there are more participants and multiple floors may arise. Our system performs real-time analysis of humans engaged in multi-party talk, which requires analyzing the distribution of pause duration and of simultaneous speech, particularly when multiple floors arise.

**CONCLUSIONS**

In this paper, we described a prototype audio space system that facilitates multiple simultaneous conversations by identifying and then separating conversational floors. We presented results from an applied conversation analytic study. The findings of user behavior and preferences have relevance for designers of audio communication systems, and the discussion of turn organization and repair strategies in the two audio spaces contributes to conversation analytic research as well. The results indicate that such a system can be quite effective in supporting parallel conversations in a natural way.

We have several immediate directions for future work. First, we will continue improving the floor assignment subsystem by using more sophisticated learning algorithms and incorporating additional features (e.g., wordspotting and intonation contours). At that point, a more formal technical evaluation of the learner accuracy will be appropriate. Second, we plan to conduct a longer-term use study of the system.

On a more basic level, we believe that automatic techniques of the kind reported here can form the basis for the lightweight, attentionally undemanding interfaces needed to support "always on," social mobile audio communication.